\documentclass[12pt]{article}
\usepackage{authblk}
\usepackage{anysize}
\usepackage{graphicx}
\usepackage{hyperref}
\usepackage{longtable}

\newcommand{\iitk}{IIT Kanpur}
\newcommand{\msft}{Microsoft Redmond}
\newcommand{\keywords}[1]{{\bf Keywords:} #1}

\newcommand{\cmt}[1]{} 
\begin{document}
    
    
    \title{Prutor: A System for Tutoring CS1 and Collecting Student Programs for Analysis}
    
    \author[1]{\mbox{Rajdeep Das, Umair Z. Ahmed, Amey Karkare\thanks{Corresponding Author}}}
    \author[2]{Sumit Gulwani}
    \affil[1]{\iitk\\\{rajdeepd,umair,karkare\}@cse.iitk.ac.in}
    \affil[2]{\msft\\sumitg@microsoft.com}
    
    \maketitle
    
    \begin{abstract}
An introductory programming course (CS1) is an integral part of any undergraduate curriculum.
Due to large number and diverse programming background of students, providing timely and personalised feedback to individual students is a challenging task for any CS1 instructor. The help provided by teaching assistants (typically senior students) is not sufficient as it suffers from unintentional bias and, most of the time, not quick enough.

In this paper, we present Prutor, a tutoring system platform to conduct introductory programming courses. Prutor is a cloud-based web application that provides instant and useful feedback to students while solving programming problems. Prutor stores, at regular intervals,  the snapshots of students' attempts to solve programming problems. These intermediate versions of the student programs provide the instructors (and data analysts) a view of the students' approach to solving programming problems.  Since Prutor is accessible through any standard web browser, students do not need to worry about dependencies external to the programming course, viz. Operating Systems, Editors, Compilers, Compiler Options, etc.. This enables the students to focus on solving only the programming problems. Apart from the code snapshots at regular intervals, Prutor also collects other valuable data such as the time taken by the students to solve the problems, the number of compile and execution events, and the errors made.  We have used this data in developing intelligent tools for giving feedback to students, some of which are described briefly in this paper. This system thus serves as a platform for tutoring as well as data collection for researchers.
\end{abstract}
    
    
    \section{Introduction}
Advancement in technology has made it possible for the computers to make education available to masses. Massive open online course (MOOC) platforms are prominent examples that have been primarily built to enhance reachability to student distributed across the world.  The successful examples of MOOCs include Coursera~\cite{coursera}, EdX~\cite{edx}, Khan Academy~\cite{khan-academy}, Udacity~\cite{udacity} and NPTEL~\cite{nptel} that are providing quality education to hundreds of thousands of students every year. These online platforms host a large number of courses, conducted by instructors from prominent institutes all over the world. These systems have also succeeded in providing open courses to the masses and improving community interaction while learning. However, even established MOOCs are lacking when it comes to providing early and useful feedback to the students during problem-solving. Problem-solving forms an integral part of learning in courses, and timely (and possibly personalised) feedback to students when solving problems can improve learning experience [CITE?].  Moreover, these MOOCs are quite generic in nature and do not address the specific challenges of conducting individual courses, for example, the dependency of a programming course on the programming environment such as the operating system, the compiler and the run-time system. These issues limit the effectiveness of online courses.

An introductory programming course (CS1) is an integral part of any undergraduate curriculum. Teaching CS1 is a challenging task due to large number and diverse programming background of students.  Since the programming assignments for the course are typically done on a computer, students expect immediate and personalised feedback from the system. Further, the online nature of the assignments makes it amenable to plagiarism.  These issues increase the workload for any CS1 instructor. The help provided by teaching assistants (typically senior students) is not sufficient because: (a) it suffers from unintentional bias (different TAs have different programming capabilities and grade similar submissions very differently based on their expertise), and (b) TAs take time to understand the (incorrect) programs and thus their feedback (or grading) is not quick enough most of the time.

Todays novice programmers need not learn the nitty-gritty of operating systems and command line compilation due to the availability of easy to use Integrated Development Environments (IDEs) having uniform interface across operating systems. However, students using an IDE typically submit the final version of their code. Thus, it is not possible to understand the approach a student used to solve the problem, the mistakes she committed, and the fixes she made. From an instructor's point of view,  it is important to {\em see} the intermediate versions of students' programs to understand the common errors committed and to provide better feedback to the student when she commits these errors.

In this paper, we present {\em Prutor (PRogramming tUTOR)}, a web based tutoring framework for teaching introductory programming. In brief, it provides an IDE for students to write and test their code for programming problems, along with interfaces to provide feedback to students on their code. For the instructors, it provides interfaces to view the code progression of student programs along with visualizations that show how the students attempt to solve the programming problems. It also helps to identify the weaker students in the class. As of now, it has been used to teach introductory C programming to the students of IIT Kanpur. The framework has been designed in a way that allows usage of other programming languages as well.

\cmt{
\begin{itemize}
\item why another online compiler?

\item an interesting  screen-shot (what is interesting?)
\item usage scenarios
\end{itemize}
\subsection{Related Work}
\subsection{Contributions}
\subsection{Organization}
}
    
    
\section{Installing Prutor}\label{sec:install}
\subsection{Requirements}\label{sec:requirements}

\begin{itemize}
\item
  Linux (any standard flavor)
\item
  Docker (\url{https://www.docker.com/})
\item
  Git (\url{https://git-scm.com/})
\item
  Python
\end{itemize}

The following are needed for maintenance and automation of certain tasks in Prutor:

\begin{itemize}
\item
  mysql client
\item
  php-mysql
\end{itemize}

\subsection{Getting Started}\label{sec:getting-started}

\subsubsection{{Step 1: Configure the installation.}}\label{step-1-configure-the-installation.}

The following are the important configurable options (and the files that
contain them):

\begin{itemize}
\item
  Number of web-application/engine containers (\_config/config.ini):
  This can be configured by editing the ``NUM\_WEBAPP'' and
  ``NUM\_ENGINE'' values inside the ``config.ini'' file in the
  ``\_config" directory.
\item
  Local codebase repositories (\_config/config.ini): This can be
  configured by modifying the ssh URLs of the ``WEBAPP\_SRC'' and
  ``ENGINE\_SRC'' parameters, located inside the ``config.ini'' file.
  Also the corresponding ``REPO\_HOST'' value must also be changed.
\item
  Components whose sources should be updated (\_config/config.ini): In
  case you updated the sources.list file mentioned above, you may want
  to restrict the use of this sources.list file for specific components.
  Setting the values of ``CACHE'', ``WEBAPP'', etc. to 1, forces the
  installer to use the custom sources.list file. A value of 0 results in
  the respective component using the default sources.list file.
\item
  APPORT (\_config/config.ini): The port where main application runs.
\item
  COURSE\_ID (\_config/env.ini): The identification number of the course
  you plan to teach
\item
  Database users (\_config/env.ini): The credentials to access mysql and
  mongo databases.
\item
  Programming Languages (nosql/init\_mongo.js): Update the list of
  programming languages and related tools for the mongodb environment.
  See the comments and existing entries in the file.
\item
  Linux repos (sources.list): In case you have a local mirror or one
  that is closer than the repository mentioned in the default
  sources.list file, you can update the sources.list file to point to
  the respective mirrors. Please take care to add a mirror containing
  all recent updates. Modify the sources.list file inside the
  ``\_config" folder to suit your needs.
\item
  Key-pairs for accessing the repositories (id\_rsa, id\_rsa.pub): The
  public/private key-pairs are used to access the codebase from the
  repositories. Update these to match the repositories that you use.
\end{itemize}

\subsubsection{{Step 2: Build the images.}}\label{step-2-build-the-images.}

Assuming that you have a fresh Linux installation with Docker and Git,
you need to first build the images corresponding to the system
components. Doing this is a breeze. All you have to do is run the
``build'' script from inside of the directory containing it. Simply run:

\begin{verbatim}
./build
\end{verbatim}

This process will take several minutes, as the individual images will
take time to build. After this is complete (assuming that there are no
errors), you should have a set of images ready. You can view these
images by typing:

\begin{verbatim}
docker images
\end{verbatim}

You should see a set of images starting with ``prutor/''. These are the
images corresponding to the various components of the system.

\subsubsection{{Step 3: Deploy and Start Prutor.}}\label{step-3-deploy-and-start-prutor.}

After the images have been built, you can start the system by running
another utility script. Simply run the following in the same directory:

\begin{verbatim}
./deploy
\end{verbatim}

This will start all the required containers and do the necessary
configurations. Once the system boots up, you shall be able to view the
following web interfaces (the ports may differ if you modified the
values in \_config/config.ini):

\begin{itemize}
\item
  Service discovery UI: http://:81/
\item
  Database Management UI: http://:84/
\item
  Web Application UI: http://:82/
\end{itemize}

The credentials for accessing the database management and service
discovery interface are given in \_config/env.ini.  It is {\em highly recommended that default credentials are replaced by secure credentials} to avoid unauthorized access to the system.

In case you encountered errors during the process, you can revert back by running the clean script. It will undo all the changes that were done by the ``build'' and ``deploy'' scripts. Just run the following to do a complete cleanup:

\begin{verbatim}
./clean
\end{verbatim}

\subsubsection{{Step 4: Enable escompiler.}}\label{step-4-enable-escompiler.}

You need to integrate escompiler feedback tool to get error messages
properly, as part of Editor UI. To do so, run the following:

\begin{verbatim}
#!shell

./_bin/integrate-feedback-tool.py \
    escompiler/escompiler.json
\end{verbatim}

\subsection{Integrating Feedback
Tools}\label{integrating-feedback-tools}

It is quite likely that you desire integrating feedback tools into the
ITS system. The task of doing so is pretty simple. You first need to
build and run the Docker containers corresponding to the feedback tools.
In the event that the containers require database or cache services, you
can specify these services using the host addresses of the respective
service. This information can be found using the service discovery user
interface. After the feedback services are up and running inside
containers, you need to make the ITS aware of them. This is the part
where you would integrate the feedback tool into the ITS. Inside the
``\_bin" directory is an executable named
``integrate-feedback-tool.py''. In order to run this script you must
first create a configuration file which contains the details of the
integration. The configuration file should be a JSON file having the
following format:

\begin{verbatim}
{
 "name": "<name_of_the_feedback_tool>",
 "services": [<services_provided_by_this_tool>]
}
\end{verbatim}

Each service provided by the tool should have the following format:

\begin{verbatim}
{
  "trigger": 
      "<event_to_execute_this_service>",
  "containers": 
      [<container_names_hosting_the_service>],
  "port": 
      <container_port_where_the_service_is_hosted>,
  "endpoint": "<url_endpoint_for_this_service>"
}
\end{verbatim}

You can refer to the ``example.json'' file in the same directory to get
an idea of how a configuration looks like. Valid values for trigger are
``ON\_COMPILE'', ``PRE\_COMPILE'', ``ON\_EVALUATE'', ``ON\_ACCEPTED'',
etc.

After creating this configuration file, invoke the above mentioned
executable in the following manner:

\begin{verbatim}
./_bin/integrate-feedback-tool.py \
     <your_configuration_file>.json
\end{verbatim}

The tool will then be integrated into the system.

\subsection{Updating}\label{updating}

You can update the codebase of the web-application and engine components
from their source repositories. In order to do so, you need to run the
``update-webapp'' and ``update-engine'' scripts, inside the ``\_bin"
directory. The web application will be restarted, once it is updated.
You can also restart the web application services by executing the
``restart-webapp'' script in the same directory.

\subsection{Scaling}\label{scaling}

You can add/remove web-application and engine containers to/from the
system in a breeze. To add a web application container to the system,
simply run the ``upscale-webapp'' script from inside the ``\_bin"
directory. Likewise, to remove a web application container invoke the
``downscale-webapp'' script from the same directory. Similar scripts
exist for the engine as well and are named ``upscale-engine'' and
``downscale-engine'' respectively.

\cmt{
\subsection{Pluggable Tools}
Contains details about tools that we could plug-in.

\subsection{Analytics}
Contains the data analytics obtained on running the tool.
}
    
    \section{Roles in Prutor}
\label{sec:roles}

\begin{figure}[t]
  \centering
  \includegraphics[width=\textwidth]{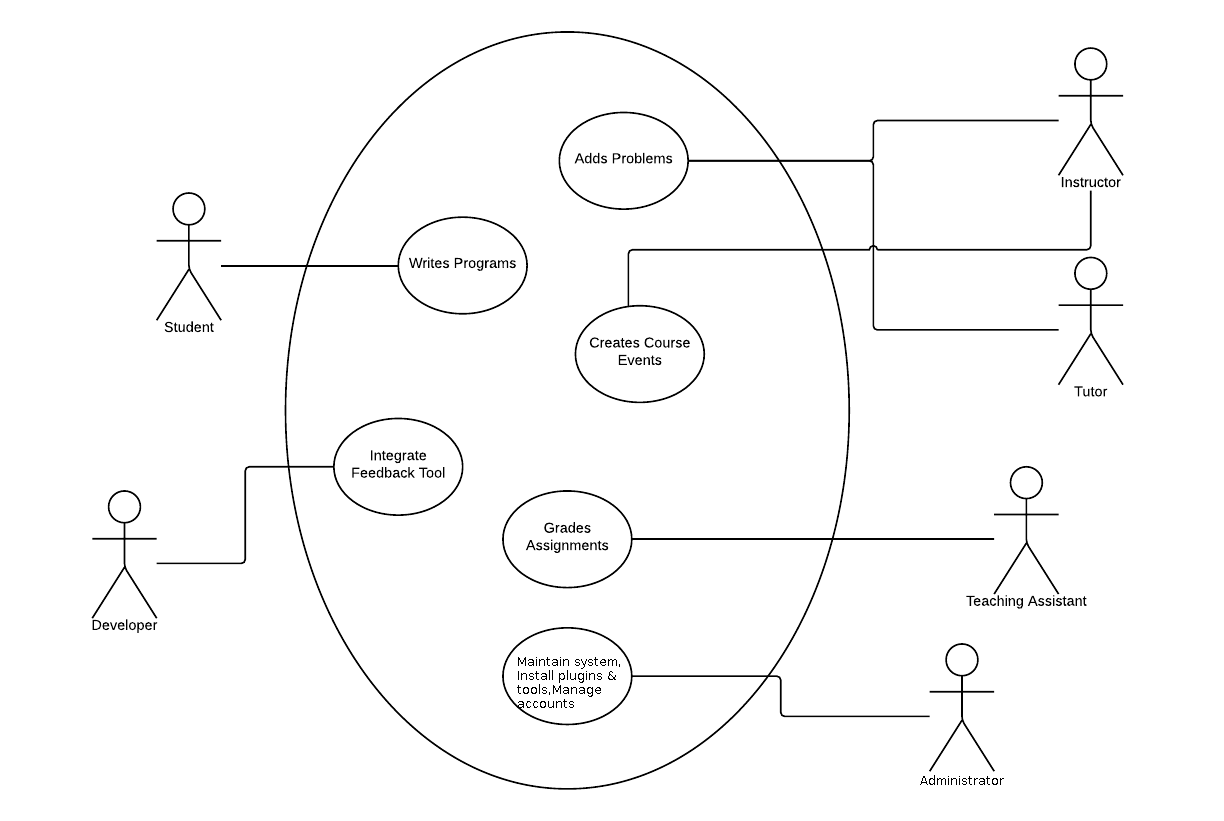}
  \caption{Roles in Prutor}
  \label{fig:logical_roles}
\end{figure}

Prutor supports the following types of roles for the users:
\begin{enumerate}
    \item Student
    \item Teaching Assistant
    \item Tutor
    \item Teacher
    \item Administrator (not available in web interface)
\end{enumerate}
Figure~\ref{fig:logical_roles} gives a logical view of the roles in Prutor. The introductory programming course at IIT Kanpur typically has 1 administrator, 1 instructor, 12 tutors, 48 teaching assistants, and approximately 450 students.

\subsection{Student}
 A student is a novice programmer who uses Prutor to learn programming. Her main task in Prutor is to submit code (graded/ungraded assignments, practice programs) for practice as well as evaluation. Prutor provides feedback to the student in case her program has syntax, semantic, run-time and/or logical errors. The amount and quality of feedback is determined by the feedback tools employed by the Prutor installation.
 
\subsection{Teaching Assistant}
 A teaching assistant (TA) helps in evaluation of students' submission by providing test-cases for problems and manually evaluating the submissions. This role has the following privileges:
\begin{itemize}
    \item {\bf Problem Management:} TAs can view problem statements, model solution to the problem and existing test-cases. They are allowed to add more test-case and delete existing test-cases.
    \item {\bf Submissions Viewer:} TAs are allowed to view the attempts made by the students for any event. 
    \item {\bf Tasks Viewer:} TAs can view the grading tasks assigned to them.
    \item {\bf Pager:} TAs can view messages and help requests posted by the students and respond to them.
    \item {\bf Code Viewer:} TAs can view the code submitted for any assignment. They can read the code, execute it for some input, and evaluate it for the existing test-cases. They can not, however, modify the submission. 
    \item {\bf Admin Editor:}  Sometime some  TA wants to modify a student's submission, say to understand the flow of some non-working part of the code in order to give partial credits.  TA can do it in the admin editor that creates a temporary copy of the code in a new browser-tab. Admin editor is like the Code Viewer, but allows the TA to modify the (copied) code. The copy is destroyed when the tab is closed.
\end{itemize}

\subsection{Tutor}
A tutor is a more trustworthy form of a TA. Apart from the privileges that a  TA has, a tutor has following extra privileges:
\begin{itemize}
    \item {Problem Upload:} A tutor can create new problems (problem statement and model solution) and update existing problems. 
\end{itemize}
Further, for highly sensitive events (called EXAMS in Prutor), the privileges of the TAs are revoked until the event is over. The tutors, however, enjoy uninterrupted access to all events.

\subsection{Teacher}
The teacher is the main role for the Prutor, typically performed by the instructor of the course. It has the following extra privileges over a tutor's privileges:  
\begin{itemize}
    \item {\bf Accounts Management:} Teacher can add other teachers, tutors, TAs and students to the course.
    \item {\bf Events Management:} Teacher can schedule LAB and EXAM events, and assign problems to these events.
    \item {\bf Control Panel:} Teacher can customize and control certain aspects of the Prutor installation, viz. the compiler used, the flags passed to the compiler, and the feedback tools deployed.
\end{itemize}
    
\subsection{Administrator}\label{sec:admin}
Administrator's main task is to manage and maintain the system. This role is not accessible from Prutor's web interface, but is available through command line on the Prutor server (a Linux system).

\section{Interfaces in Prutor}
\label{sec:interfaces}
Prutor has several interfaces to cater to the needs of different roles played by the users. We describe these interfaces in details next.

\subsection{Student Home}\label{sec:student-home}
This page displays the following information to students:

\begin{enumerate}    
\item {\bf View grades and course statistics:} This page allows student users to view their grades for previous course assignments. This also enables them to view their own statistics for the course, such as the number of submitted solutions. 
\item{\bf View ongoing events:} The home page for students displays any ongoing course event along with a summary of the problems that have been assigned to the student for the respective course event. A student can directly go to the environment from where he can start coding for any of the problems listed in the summary.
\end{enumerate}

\subsection{Assignment Editor}\label{sec:assign-editor}
Te assignment editor has the following services:
\begin{enumerate}
\item{\bf View Problem Statement:} This interface displays a programming problem that has been assigned to the respective student for the current ongoing course event. The student is able to view this problem from one of the tabs located to the left of the user interface.
\item{\bf Edit Program:} This service allows student users to write source code for the programming problem that is defined within this interface. Code can be written in an editor that is displayed within this interface. The editor support basic features such as syntax highlighting, code-folding, automatic indentation and displaying line numbers. Students are also able to save their programs manually.
\item{\bf Compile Program:} This service allows student users to compile the programs that they have written in the editor. They are also able to view the compiler messages in a console window within this interface, along with any feedback that are provided by the feedback tools integrated into this system. 
\item{\bf Execute Program:} This service allows users to execute compiled programs that they have written in the editor. They can supply input data into a corresponding window located within this interface. The output from the program is displayed in another similar window located adjacently. This interface also displays any error messages that may have occurred while executing the program.
\item{\bf Evaluate Program:} This service allows users to evaluate their program against a set of test cases supplied along with the problem statement. They are able to view the number of passed test cases along with a table showing which have passed and which have failed, if the respective test cases are visible. The interface also displays any feedback generated by any of the relevant feedback tools integrated into this system.
\item {\bf Submit Program:} This service allows student users to submit their programs as solutions for the current programming problem. 
\end{enumerate}

\subsection{Scratchpad}\label{sec:scratchpad}
\begin{enumerate}
\item {\bf Create/Delete files/folders:} This service allows users to create and delete virtual source files, and folders to organize them.
\item{\bf Write Programs:} This service allows users to write arbitrary programs for practice or testing purposes. An editor is provided within this interface, which is similar to the one present in the editor for course events. These programs are saved in the virtual source files created above.
\item{\bf Compile:} This service allows users to compile their source code, written in the virtual source files, and view the messages generated by the compiler.
\item{\bf Execute:} This service allows users to execute their compiled source code on custom input, and view the results in a corresponding output window. Users are also able to view any errors which occur while executing the program.
\end{enumerate}

\subsection{Codebook}\label{sec:codebook}
\begin{enumerate}
\item{\bf View attempted problems list:} This service allows users to view the list of practice problems that they have attempted, and also the list of problems they had been assigned during the course events. Students can view these problems and their own solutions to them.
\item{\bf View submitted solution:} This service allows students to view the code submitted as solution for the problems listed within this interface. For practice problems, this interface displays the last saved solution code. It also displays the grading status of the problem, if the problem is an assignment problem. 
\end{enumerate}

\subsection{Practice arena}\label{sec:practice}
\begin{enumerate}
\item{\bf View practice problems:} This service allows student users to view the problems that have been marked for practice. Students can directly navigate to the “Editor for course events” interface from here, for the corresponding problem.
\end{enumerate}

\subsection{Admin home}\label{sec:admin-home}
\begin{enumerate}
\item {\bf Navigate to management and analytics interfaces:} This interface is used by administrators to navigate to the management and data analytics interfaces. It provides an overview of what features are available to the user.
\end{enumerate}

\subsection{Problem management}\label{sec:problem-mgmt}
\begin{enumerate}
\item{\bf Create problems:} This service is used to create a new blank problem instance by specifying an identifier for the problem, along with a category that the problem belongs to.
\item{\bf View/Edit problems:} This service is used to view and edit previously created problems. Items that can be edited for a problem are its statement, solution code and solution template. All edits are reflected instantly to anyone viewing the respective problem. 
\item{\bf Add/Delete test cases:} This service allows instructors, tutors and teaching assistants to add test cases for the problem, upon which student solutions can be evaluated. Test cases can be added individually, or in batches. Test cases can also be deleted using this interface.
\end{enumerate}

\subsection{Problem upload}\label{sec:problem-upload}
\begin{enumerate}
\item{\bf Upload problems in batch:} This service allows problems to be uploaded to the system in batch. The problems are required to be in a specific format, in order to be uploaded.
\end{enumerate}

\subsection{Account management}\label{sec:account}
\begin{enumerate}
\item{\bf Add accounts:} This service allows administrators to add user accounts to the system. User accounts may be either student or administrator accounts.
\item{\bf Edit/Delete accounts:} This service allows administrators to edit profile and accounting information of existing users, including authentication methods. It also allows deleting of user accounts from the system. For administrator accounts, this interface also facilitates changing of administrator roles such as instructor, tutor and teaching assistant.  
\end{enumerate}

\subsection{Event management}\label{sec:event-mgmt}
\begin{enumerate}
\item{\bf View course event calendar:} This service allows viewing of course events and their corresponding schedules. The interface includes a calendar, which helps instructors to plan course events.
\item{\bf Create course events:} This service allows the creation of new course events in the system, such as labs, exams and quizzes.
\item{\bf Create event schedules:} This service allows instructors to create schedules for the course events which had been created previously.
\item{\bf Assign problems to students:} This service allows instructors to assign problems to students for course events that had been created previously. This assignment is done based on some algorithm, which is interactively specified by the instructor.
\end{enumerate}

\subsection{Event dashboard}\label{sec:event-dash}
\begin{enumerate}
\item{\bf View student performance rankings:} This service allows instructors to view rankings of students according to some score computed using some algorithm. 
\item{\bf View student performance distributions:} This service allows instructors to view the distribution of the scores mentioned above.
\end{enumerate}

\subsection{Submissions viewer}\label{sec:submissions}
\begin{enumerate}
\item{\bf View student submissions:} This service enables viewing of student submissions to problems for course events. The submissions can be filtered based on a set of categories.
\end{enumerate}

\subsection{Code history viewer}\label{sec:code}
\begin{enumerate}
\item{\bf View code history:} This service allows viewing of source code history for a specific course assignment. It is used to view the stimuli on which code versions were saved along with the code that was saved. 
\item{\bf View submission code:} This service allows viewing of the code version that was submitted by the student, while solving this assignment.
\item{\bf Evaluate submitted code:} This service enables evaluation of the submitted source code on the test cases that are available for the respective problem. Evaluation results for each test cases are displayed in a table.
\item{\bf Grade submissions:} This service enables teaching
  assistants and tutors to grade the assignment being viewed.
\end{enumerate}

\subsection{Admin editor}\label{sec:admin-editor}
\begin{enumerate}
\item{\bf Compile program:} This service allows admin users to compile programs and view syntactic error messages.
\item{\bf Execute program:} This service allows admin users to execute programs on custom input.
\item{\bf Evaluate program:} This service allows admin users to evaluate programs that correspond to a student submission for a course event. Admins are able to modify the code and then evaluate on the modified code.
\item{\bf Update solution code:} This service allows admin users to update the ground truth solution code to a particular problem.  
\end{enumerate}

\subsection{Assignment analytics}\label{sec:assignment}
\begin{enumerate}
\item{\bf View code size variation:} This service allows users to view the variations in code size of a student submission over time.
\item{\bf View code save progression:} This service allows users to view the progression of code saves for a student submission, over time.
\item{\bf View syntactic analytics:} This service allows users to  view the sort of compilation errors performed by a student while solving that particular programming assignment. It also shows instances of the errors and where they were triggered.
\item{\bf View compilation error timeline:} This service allows users to view the compilation errors committed by the students in solving the problem, as well as the time taken by them to fix the individual compilation errors.
\item{\bf View execution sequence:} This service allows users to view the sequence of executions and evaluations made by a student while solving a programming assignment, over time. It also displays the results of those executions and evaluations.
\end{enumerate}

\subsection{Control panel}\label{sec:control}
\begin{enumerate}
\item{\bf Modify compiler options:} This service allows admins to modify the flags used by the compiler to compile programs.
\item{\bf Modify execution sandbox settings:} This service allows admins to change the quotas for execution of programs. Quotas include time and memory.
\item{\bf Enable/Disable plugins:} This service allows admins to enable or disable plugins used in the system.
\item{\bf Modify delays:} This service allows admins to modify the time delays for compilation, execution and evaluation of programs.
\item{\bf Enable/Disable logging:} This service allows admins to enable or disable logging of compilations, executions and evaluation attempts by students, while solving programming problems.
\end{enumerate}

\subsection{Tasks panel}\label{sec:tasks}
\begin{enumerate}
\item{\bf View personal pending/complete tasks:} This service allows admin users to view the pending or complete tasks, which have been assigned to them.
\item{\bf View overall pending/complete tasks:} This service allows admin users to view the status of tasks that have been assigned to all admin users, grouped by the admin users and the course events for which tasks have been assigned.
\end{enumerate}

\subsection{Admin account settings}\label{sec:admin-account}
\begin{enumerate}
\item{\bf Update name:} This service allows admin users to update their name on the system.
\item{\bf Update password:} This service allows admin users to update their password on the system.
\end{enumerate}

\subsection{Consul interface}\label{sec:consul}
\begin{enumerate}
\item{\bf View health status of nodes:} This service allows users to view the health status of the various nodes deployed in the system.
\end{enumerate}

\subsection{PhpMyAdmin  interface}\label{sec:phpmyadmin}
\begin{enumerate}
\item{\bf View health status of nodes:} This service allows users to view the health status of the various nodes deployed in the system.
\end{enumerate}

    \begin{figure}[t]
\centering
\includegraphics[width=\textwidth]{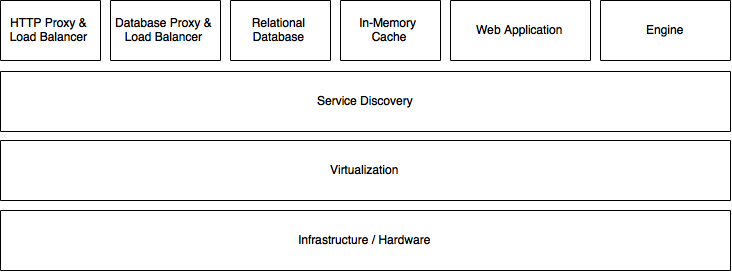}
\caption{A logical view of the system}
\label{fig:logical_view}
\end{figure}

\section{Architecture of Prutor}\label{sec:arch}
The system consists of a set of components each of which are responsible for delivering specific services. Some of these services are exposed to the client user, whereas the remaining are exposed to the other services. The components can be plugged into and out of the system dynamically, making the system easily scalable. A service discovery layer exists below these components enabling them to know of each other. Feedback tools are integrated into the system as similar components with restricted access to the services offered by the system. Figure~\ref{fig:logical_view} shows a logical view of the system.

Technically speaking, the system is ``Dockerized''. Every component is a Docker~\cite{docker} container, knit together using the Docker network. Components are added to the system by spawning new Docker containers and removed by deleting them. Service discovery is achieved by running an agent in every ``known'' Docker container and providing them with the locations of any one of the running containers within the network. The feedback tools form the ``unknown'' containers whose functionality is not known to the other components. Thus, these containers need to be explicitly provided links to the services that they require (within the network).

The use of Docker containers enables portability across platforms. Due to its lightweight nature, a minimal overhead is incurred in running the components. Further, spawning new containers takes time of the order of seconds, which makes it possible to write scripts for scaling the system automatically.

The tutoring system has 6 principal components (Figure~\ref{fig:logical_view}). These are 
\begin{enumerate}
\item HTTP proxy, 
\item Database proxy, 
\item Relational database, 
\item In-memory cache, 
\item Web application, and 
\item Engine.
\end{enumerate}
All these components reside on top of a service discovery layer, which is responsible for keeping track of the health and presence of the instances of these components. Below the service discovery layer lies a virtualization layer, which enables creation and deletion of instances dynamically within a physical machine. It also enables distribution and relocation of instances across physical machines, without any changes to the instances themselves. The physical hardware forms the lowest layer of the architecture.

The system is designed to scale horizontally. It can handle the addition and removal of service instances dynamically. This requires keeping track of what services are running and where they are running. Also components within the system, which need to interact with one another need to know of each other in a dynamic environment. Thus, service discovery is used in the deployment of the system. Consul~\cite{consul} acts as the service discovery agent in the various components across the system. Whenever a component boots up, it adds itself to the system cloud, by registering itself as a service on the system. Other components, which require the services offered by this component consults the service discovery agent to get the location and configuration of the service providing node. It then uses this information to communicate with the node. Whenever a component leaves the system, it de-registers itself from the service cloud and exits. This way, nodes can be dynamically added and removed from the system without affecting service availability. Another important feature of the service discovery agents is to keep track of the health of the individual nodes. By registering checks for itself on its agent, each node publishes its health status via the agent to the service cloud. This way, monitoring applications can keep track of what services are unavailable and should be replaced or repaired.

\subsection{HTTP Proxy}
This component acts as a proxy as well as a load balancer for all HTTP requests corresponding to the web services offered by the Engine and Web Application. Requests are redirected to either the Engine or the Web Application instances according to some URL rules specified in its access control list. It also load balances the requests among all the instances of each of these two types of web servers. The least connected load balancing algorithm is used to distribute the requests. The proxy uses service discovery to keep track of what instances of each web service are available and healthy. It uses this information to update its server list whenever there is a change. This mechanism enables instances to be dynamically added and removed from the system without any downtime.

\subsection{Database Proxy} 
This component acts as a proxy and a load balancer for the database instances. The job of this component is to redirect requests for database accesses to one of the available database instances, based on the least connected load balancing algorithm. The proxy acquires information about the available database instances through a service discovery agent. It updates its server list based on the health and availability of such instances. 

The rationale behind the requirement of a database proxy is to uniformly distribute database load across all the available database instances. It also helps to dynamically add and remove database instances from the system without any downtime. As database queries can vary based on the amount of time required to address them, it is not wise to statically assign database instances to application nodes which require it. This is because, a single application node can consume a lot of database resources, while others may be almost idle. Thus the entire load gets concentrated on a single database node, which is inefficient. A load balancer is kept to tackle this situation, which balances load among the running instances of database nodes.

\subsection{Relational Database} 
This component is responsible for storing all data corresponding to the system. The data stored includes user accounting information, code history, programming problems and their test cases, course events and corresponding assignments, grading information, compilation logs, execution logs and evaluation logs. The database also stores various other information required for system operation such as configurations. This component is the sole persistent data store for the system. The database instances are clustered using replication, so that they together form a high-availability, fault-tolerant and durable data store. The cluster uses master-master replication with synchronous writes. This implies that all database instances have the same data at any instant of time. This feature guarantees data consistency.

One of the main purposes of the system is to provide a framework for collection of data for analytics. Since the data collected by the system is relational in nature, a relational database has been chosen as the data storage engine. Also a relational database supports querying in ways which support providing statistical information directly. This is necessary for various research activities which require such information.

\subsection{In-Memory Cache} 
This component is responsible for session storage and database caching. Use of an in-memory storage for sessions ensure high performance, where almost all of the requests for web services require authentication. Database caching is also necessary as a lot of database queries are expensive and do not change frequently. The cache instances are also clustered. However the data across the instances are sharded to optimize performance. The cache component has no persistence, and in the event of a node failure, all data stored in that node is lost. However, since the data items stored in the cache instances are not valuable and are recoverable, this is not a problem. The cache cluster is not replicated to minimize the overhead involved. 

Session stores mostly use persistence, in which they save the contents to the disk in either regular files or databases. While designing the session store, it was observed that almost all of the requests that are served by either of the services - engine or web application, require authentication. Thus from the perspective of the web applications, they must lookup the session stores to fetch/write information on every request. For a session store that uses persistence, this is quite expensive and may be a bottleneck for the application itself. Thus an in-memory cache is used as a session store, to speed up the process. Memcached is used for this purpose as it can be sharded and distributed across nodes with ease. One may argue that using a persistent store with caching such as Redis can be a better option, as it has persistence as well as in-memory caching. However, firstly clustering support for Redis is not yet stable and moreover writes to Redis will cause an overhead larger than Memcached. Further, as session data is not valuable data, persistence is not necessary. Losing session data simply results in the end user being logged out. We can always add redundancy to the Memcached cluster to avoid such losses, if necessary.

\subsection{Web Application}
This component is responsible for providing user interfaces and API services corresponding to the web application offered by the system. The operations performed by this component are mostly database and hence I/O intensive. Thus a web server with asynchronous I/O is chosen for this purpose. Each instance of this component consumes a single thread, and hence multiple instances are required to increase performance. Thus, the web application subsystem is clustered as well. However each node in the cluster is independent of one another. The HTTP proxy balances the load among these nodes. The web application subsystem has the largest codebase among all the components present in the system. 

\subsection{Engine}
This component is responsible for handling all compilation, execution and evaluation requests of the tutoring system, along with requests for execution of some tools. It performs all of the major compute intensive operations of the system. The Engine uses a multi-threaded web server to handle its requests, so that each request can be assigned to a single thread. This component is also clustered so that Engine nodes can be added and removed dynamically from the system. The HTTP proxy balances the requests among these nodes.

The engine is a web application which runs behind a web server. Since the time taken for the operations that the engine performs can be quite high, it is necessary to queue them. However, an explicit message queue is not maintained for this purpose. Instead the web server is configured to receive a maximum of n clients for processing by the engine. The value of n is carefully chosen such that the instance is able to process all the n requests simultaneously. Any more requests that come in during the time when the engine is saturated, get queued by the web server automatically. Thus a queue is implicitly maintained by the web server. Moreover the timeouts for the web server are chosen such that they exceed the maximum expected engine time for any request.

The services provided to the users are split between two components - the engine and the web application server. This is done intentionally, to enable scaling of the system based on usage of services and provisioning the components on different hardware based on the type of resources that they consume most. The web application is a database intensive piece of software, which in turn implies that it is an I/O intensive application. An application server which supports this sort of usage is Node JS. Thus the web application server is built using it. On the other hand, services which involve compilation and execution are CPU intensive and require an environment which supports this. Apache coupled with PHP is used for this purpose. Although it may be a good option to use “better” environments such as Java for this purpose, this choice has been made from the perspective of rapid prototyping. Since Apache spawns new threads for each request, a separate thread can be dedicated for an engine request. Node JS on the other hand cannot support this as it is single threaded. 

\subsection{Static Content Proxy}
Apart from the regular web and application servers, additional proxy servers are maintained in order to serve static content. This has not been included in the diagram above, as it is optional. The static content proxy server is used to relieve the application servers of unnecessary load, and also to speed up site loading in browsers by strategically placing static content on high speed web servers. 

\begin{figure}[t]
\centering
\includegraphics[width=.98\textwidth]{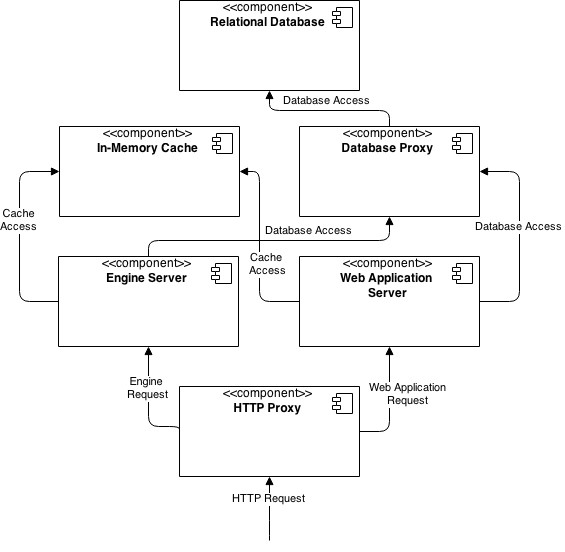}
\caption{Connectivity of the Components}
\label{fig:comp_connect}
\end{figure}

Figure~\ref{fig:comp_connect} describes the connectivity of the components described above. All HTTP requests arrive at the HTTP proxies. The proxy servers then forwards these requests to either the Web Application servers or the Engine servers, depending upon the URL path of the request. Load balancing at the HTTP level happens at this step. Each of the Engine and Web Application  servers requires the services of the Database and In-Memory Cache servers. The cache servers are accessed directly, whereas the database servers are accessed via the Database Proxy.  

\begin{figure}[t]
\centering
\includegraphics[width=\textwidth]{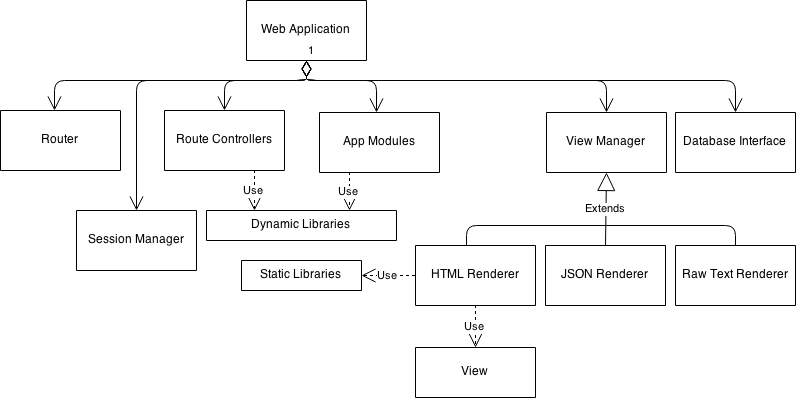}
\caption{A Logical View of the Web Application Subsystem}
\label{fig:logical_view_web}
\end{figure}

The web application is composed of four basic components - Router, Route Controllers, App Modules and the View Manager (Figure~\ref{fig:logical_view_web}). The Router is responsible for invoking the appropriate action whenever a web request comes in. Based on the HTTP request, it invokes the corresponding controller. It is also responsible for filtering out non-authentic requests. It uses the session manager to check whether the request belongs to a valid user session or not. It also checks whether the user belongs to the correct role required to access the requested resources. The controller is responsible for a group of routes which handle similar functionalities. It acts as a coordinator between the various application modules and the view manager. It invokes the required modules with the necessary arguments, based on the request and forwards the final output to the view manager for rendering to the client. The application modules handle specific sets of functions. They interact with the database of the system, using the database interface,  to perform basic CRUD operations as well as some computations. The view manager is responsible for rendering output to the client, after all necessary data retrieval and computations have been performed. The view manager renders the output in one of the three formats - HTML, JSON or raw text. The HTML renderer is generally used for rendering user interfaces, whereas the JSON and raw text renderer are used to output data requested via API calls. The HTML renderer uses view definitions to render its output. 

The web application uses a variation of the Model View Controller (MVC) architecture. Each route has a controller, which can use multiple modules in the application.

\begin{figure}[t]
\centering
\includegraphics[width=\textwidth]{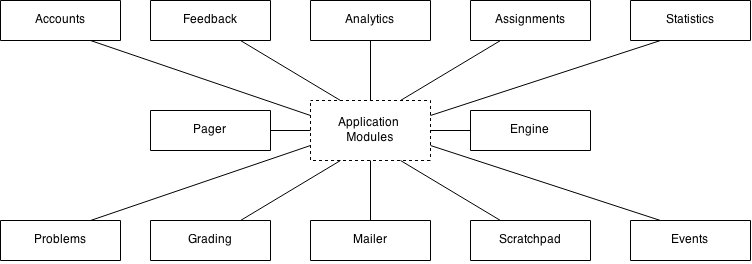}
\caption{The application modules in the Web Application subsystem}
\label{fig:web_app_view}
\end{figure}

The web application subsystem contains a number of modules which are responsible for various features of the system  (Figure~\ref{fig:web_app_view}). The accounts module is responsible for managing user accounts on the system as well as handling user authentication. The problems module is responsible for managing programming problems and their test cases used in conducting the course. The events module manages the course events and their schedules. The statistics module is used to provide statistical information regarding the conducted course. Examples of statistical information include number of submitted solutions, number of correct solutions, number of labs conducted, etc. The feedback module is responsible for managing data for feedback tools, integrated into the system. The analytics module is used for generating analytics from the data collected by the system. The assignments module is responsible for managing the course assignments and maintaining code history for the assignments. The pager handles the messaging system within the application. The engine module is responsible for updating the engine configurations to the database, so that they can be synchronized across nodes. The grading module manages the grades of assignments for various course events. The mailer module is used to send email messages by the system. The scratchpad module handles the creation, deletion and modification of files in the student scratchpad.

\begin{figure}[t]
\centering
\includegraphics[width=\textwidth]{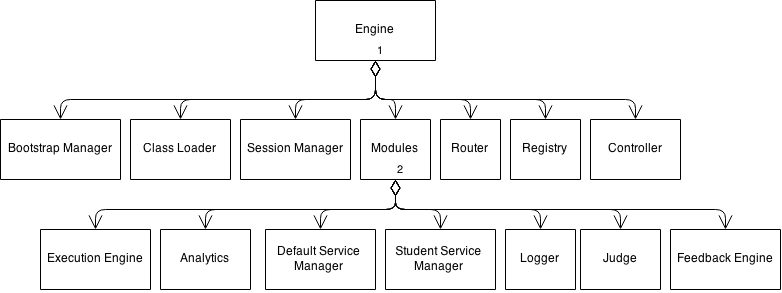}
\caption{ A logical view of the Engine subsystem}
\label{fig:engine_view}
\end{figure}

The engine is composed of six major components - bootstrap manager, session manager, application modules, router, registry and controller (Figure~\ref{fig:engine_view}) . The bootstrap manager boots the sub-system on every request, so as to include the necessary classes and methods for use in the later stages. Only the necessary classes and modules are loaded by the bootstrap manager, which includes the class loader. The class loader loads the required classes dynamically on each request. The session manager uses the cookie information provided with the request to load the user session into the request. Once the bootstrap manager finishes booting, it invokes the router with the request path and parameters. The router then decides the action to be performed based on the request path, and invokes the appropriate controller for the same. The router uses the registry to decide which controller to invoke, based on a mapping created previously. The controller coordinates between the application modules to perform the required action, as specified by the request. It creates objects of the necessary modules, once it gets the request from the router, and invokes the appropriate methods on them. It also uses the session manager to decide if the request is authentic or not and drops the non-authenticated requests with an error or redirection. The application modules are not completely decoupled, and use methods from other modules to achieve their purposes. The final result received by the controller from the application modules is then converted into a JSON string and sent back to the client. 

The modules in the engine subsystem deliver the functionality offered by it. The execution engine is responsible for compiling and executing programs. The analytics module is used to perform analysis on data collected by the system. The default service manager handles engine requests by admin users and students using the system for purposes other than course events. The student service manager handles requests by students using the system for course events. The logger logs the results of compilation, execution and evaluation for course events. The judge is responsible for evaluating student programs. The feedback engine generates feedback for the student programs. 

The engine has a variation of the Model View Controller (MVC) architecture. Here, the controller does not have a dedicated model to itself. Instead, it can communicate with multiple modules to achieve its purpose. This promotes reuse of modules within the system.

\section{Summary of Components and Interfaces}
We summarise the important components and interfaces present in Prutor.

\subsection{Components}
{\centering

\renewcommand{\arraystretch}{1.5}
\begin{tabular}{|p{.3\textwidth}|p{.65\textwidth}|} \hline
{\bf Component} & Engine \\ \hline 
{\bf Responsibilities} &
Compiles, executes and evaluates programs. It also runs various tools integrated into the system. Examples of the tools are the feedback generation tools. 
\\ \hline 
{\bf Interfaces Provided} &
Engine endpoint
\\ \hline 
{\bf Collaborators} &
\begin{tabular}[t]{|p{.3\textwidth}|l|} \hline
{\bf Interface} & {\bf Component} \\ \hline
Database proxy & 
Relational Database \\ \hline
Cache interface & 
In-Memory Cache \\ \hline  
\end{tabular} \\ &
\\ \hline
{\bf Remarks} &
Uses the database to read configurations. It is dynamically created and destroyed. It is created when a engine node is spawned. It is destroyed when the node is destroyed. A single engine instance uses multiple threads. \\ \hline 
{\bf Issues / Enhancements} & 
Handle multiple programming languages and paradigms. 
\\ \hline 
\end{tabular}
\ \\[8mm]

\begin{tabular}{|p{.3\textwidth}|p{.65\textwidth}|} \hline
{\bf Component} &
Web Application \\ \hline
{\bf Responsibilities} &
Responsible for rendering user interfaces for the tutoring system. Handles API requests corresponding to the web application, such as code saves, fetching grade cards, fetching assignment problems, etc. Also responsible for creating and destroying user sessions for the application.
\\ \hline
{\bf Interfaces Provided} &
 \begin{tabular}[t]{rl}
   1.&  Web application interface \\
   2. & User interfaces
 \end{tabular}
 \\ \hline
{\bf Collaborators} &
\begin{tabular}[t]{|p{.3\textwidth}|l|} \hline
{\bf Interface} & {\bf Component} \\ \hline
Database proxy&
Relational Database \\
Cache interface &
In-Memory Cache \\
\end{tabular}
 \\ \hline
{\bf Remarks} &
Persistent for the lifetime of the node on which it runs. Destroyed when the node is destroyed. An instance is single threaded. \\ \hline
{\bf Issues / Enhancements} &
Improve Error handling. \\ \hline
\end{tabular}
\pagebreak

\begin{tabular}{|p{.3\textwidth}|p{.65\textwidth}|} \hline
{\bf Component} &
In-Memory Cache \\ \hline
{\bf Responsibilities} &
Stores session information and database rows for caching.
\\ \hline
{\bf Interfaces Provided} &
Cache interface
 \\ \hline
{\bf Collaborators} &
\\ \hline

{\bf Remarks} &
Created when a cache node is created and destroyed when it is destroyed. \\ \hline
{\bf Issues / Enhancements} &
Sharding algorithms used by the engine and web application nodes do not match perfectly. \\ \hline
\end{tabular}
\ \\[8mm]

\begin{tabular}{|p{.3\textwidth}|p{.65\textwidth}|} \hline
{\bf Component} &
Relational Database \\ \hline
{\bf Responsibilities} &
Stores all persistent data for the system and collected by the web services. 
\\ \hline
{\bf Interfaces Provided} &
Database interface
 \\ \hline
{\bf Collaborators} &
 \\ \hline
{\bf Remarks} &
Created when a database node is created and destroyed when it is destroyed. Uses multiple threads. \\ \hline
{\bf Issues / Enhancements} &
Clustering uses synchronous writes. Hence writes are slow. \\ \hline
\end{tabular}
\ \\[8mm]

\begin{tabular}{|p{.3\textwidth}|p{.65\textwidth}|} \hline
{\bf Component} &
Database Proxy \\ \hline
{\bf Responsibilities} &
Load balances database requests among the available database nodes.

The provided interfaces are:
Database proxy interface
 \\ \hline
{\bf Collaborators} &
\begin{tabular}[t]{|p{.3\textwidth}|l|} \hline
{\bf Interface} & {\bf Component} \\ \hline
Database interface &
Relational Database \\ \hline
\end{tabular}
\\ \hline

{\bf Remarks} &
Created when a new system is deployed. Persists throughout the lifetime of the system. Destroyed only when replaced. \\ \hline
{\bf Issues / Enhancements} &
Connection timeouts need to be inferred from system performance. \\ \hline
\end{tabular}
\ \\[8mm]

\begin{tabular}{|p{.3\textwidth}|p{.65\textwidth}|} \hline
{\bf Component} &
HTTP Proxy \\ \hline
{\bf Responsibilities} &
Load balances HTTP requests among the Web Application and Engine nodes. 

The provided interfaces are:
HTTP proxy interface    
 \\ \hline
{\bf Collaborators} &

\begin{tabular}[t]{|p{.3\textwidth}|l|} \hline
{\bf Interface} & {\bf Component} \\ \hline
Web Application Interface & 
Web Application \\ \hline
Engine endpoint &
Engine \\ \hline
\end{tabular}
 \\ \hline
{\bf Remarks} &
Created when a new proxy node is created. Destroyed when it is destroyed. Usually remains throughout the lifetime of the system. \\ \hline
{\bf Issues / Enhancements} &
Load balancing weights need to be determined based on the availability of web application and engine nodes. \\ \hline
\end{tabular}
}

\subsection{Interfaces}
{ \centering

\renewcommand{\arraystretch}{1.5}
\begin{longtable}{|p{.3\textwidth}|p{.65\textwidth}|} \hline
{\bf Interface} &
Web Application \\ \hline
{\bf Description} &
Provides access to the Web Application services. \\ \hline
{\bf Services} &
\begin{tabular}{@{}p{.6\textwidth}}
{\bf Operation:} render login page\\
{\bf Description:} Renders the user interface to enable users to log into the system.\\ \\

{\bf Operation:} login \\
{\bf Description:} Attempts to login a user to the system. \\ \\
\end{tabular} 
\\ {\bf Services contd\ldots  (Web Application Interface)} & \begin{tabular}{@{}p{.6\textwidth}}
{\bf Operation:} logout \\
{\bf Description:} Logs out an user from the system and clears session information. \\ \\

{\bf Operation:} render home page \\
{\bf Description:} Renders the home page for a student. \\ \\

{\bf Operation:} render event editor interface \\
{\bf Description:} Renders the editor interface corresponding to a course event. \\ \\

{\bf Operation:} render scratchpad interface \\
{\bf Description:} Renders the editor interface for the student scratchpad.  \\ \\

{\bf Operation:} render practice arena \\
{\bf Description:} Renders the user interface containing practice problems. \\ \\

{\bf Operation:} render codebook \\
{\bf Description:} Renders the user interface corresponding to a student and containing all the code submissions for course events as well as practice problems. \\ \\

{\bf Operation:} render codebook page \\
{\bf Description:} Renders the user interface containing a code view, grading information and problem statement corresponding to a student’s codebook entry. \\ \\

\end{tabular} 
\\ {\bf Services contd\ldots  (Web Application Interface)} & \begin{tabular}{@{}p{.6\textwidth}}
{\bf Operation:} save code for assignment \\
{\bf Description:} Saves a code version for an assignment or practice problem, in  a certain mode.  \\ \\

{\bf Operation:} create pager message \\
{\bf Description:} Creates a new message on the pager feature, corresponding to a certain context.  \\ \\

{\bf Operation:} respond to pager message \\
{\bf Description:} Creates a response for a pager message thread. \\ \\

{\bf Operation:} delete pager message \\
{\bf Description:} Deletes a message corresponding to a pager thread. \\ \\

{\bf Operation:} render pager view \\
{\bf Description:} Renders the user interface containing all message threads created by the student. \\ \\

{\bf Operation:} create a file \\
{\bf Description:} Creates a virtual scratchpad file on the system. \\ \\

{\bf Operation:} delete a file \\
{\bf Description:} Deletes a scratchpad file from the system. \\ \\

{\bf Operation:} save scratchpad file \\
{\bf Description:} Saves the contents of a scratchpad file to the database. \\ \\
\end{tabular} 
\\ {\bf Services contd\ldots  (Web Application Interface)} & \begin{tabular}{@{}p{.6\textwidth}}

{\bf Operation:} create a folder \\
{\bf Description:} Creates a virtual ScratchPad folder on the system. \\ \\

{\bf Operation:} render admin home \\
{\bf Description:} Renders the admin home user interface, from where various admin services can be accessed. \\ \\

{\bf Operation:} render user accounts \\
{\bf Description:} Renders the user interface listing the user accounts on the system. \\ \\

{\bf Operation:} create admin account \\
{\bf Description:} Creates a new admin user account on the system. \\ \\

{\bf Operation:} create student account \\
{\bf Description:} Creates a new student user account on the system. \\ \\

{\bf Operation:} delete admin account \\
{\bf Description:} Deletes an admin user account from the system. \\ \\

{\bf Operation:} delete student account \\
{\bf Description:} Deletes a student user account from the system. \\ \\

\end{tabular} 
\\ {\bf Services contd\ldots  (Web Application Interface)} & \begin{tabular}{@{}p{.6\textwidth}}
{\bf Operation:} modify admin role \\
{\bf Description:} Modifies the role taken by an admin on the system, such as an instructor, tutor or teaching assistant. \\ \\

{\bf Operation:} modify admin name \\
{\bf Description:} Modifies the name by which the admin user is identified on the system. \\ \\

{\bf Operation:} modify admin password \\
{\bf Description:} Modifies the password for an admin user. \\ \\

{\bf Operation:} modify student account \\
{\bf Description:} Modifies profile and accounting information related to a student user. \\ \\

{\bf Operation:} render problem management portal \\
{\bf Description:} Renders the user interface for managing problems for the course which is being run using the system. \\ \\

{\bf Operation:} render problem view \\
{\bf Description:} Renders an interface containing all information related to a problem, and where the problem can be edited.  \\ \\

{\bf Operation:} render problem upload \\
{\bf Description:} Renders the interface using which problems can be uploaded to the system in batch. \\ \\

\end{tabular} 
\\ {\bf Services contd\ldots  (Web Application Interface)} & \begin{tabular}{@{}p{.6\textwidth}}

{\bf Operation:} update problem statement \\
{\bf Description:} Updates the problem statement for a problem instance. \\ \\

{\bf Operation:} update problem solution \\
{\bf Description:} Updates the solution code for a problem. \\ \\

{\bf Operation:} update problem template \\
{\bf Description:} Updates the initial template corresponding to a problem. \\ \\

{\bf Operation:} update problem title \\
{\bf Description:} Updates the title corresponding to a problem. \\ \\

{\bf Operation:} add test case \\
{\bf Description:} Adds a test case for a specific problem. \\ \\

{\bf Operation:} remove test case \\
{\bf Description:} Removes a test case from the system. \\ \\

{\bf Operation:} add bulk test cases \\
{\bf Description:} Adds test cases for a problem in batch. \\ \\

{\bf Operation:} delete problem \\
{\bf Description:} Deletes a problem from the system. \\ \\
\end{tabular} 
\\ {\bf Services contd\ldots  (Web Application Interface)} & \begin{tabular}{@{}p{.6\textwidth}}

{\bf Operation:} mark problem practice \\
{\bf Description:} Marks a specific problem as a practice problem. \\ \\

{\bf Operation:} render event management portal \\
{\bf Description:} Renders the user interface using which course events can be added, schedules can be made and problems can be assigned to students. \\ \\

{\bf Operation:} create event \\
{\bf Description:} Creates a course event on the system for the ongoing course. \\ \\

{\bf Operation:} delete event \\
{\bf Description:} Deletes a previously created event from the system. \\ \\

{\bf Operation:} schedule event \\
{\bf Description:} Creates a schedule for a particular event. \\ \\

{\bf Operation:} assign problems \\
{\bf Description:} Assign problems for a course event to students based on some algorithm. \\ \\

{\bf Operation:} add slots for schedule \\
{\bf Description:} Add slots for a specific schedule of an event. Slots define which student correspond to a particular schedule. \\ \\
\end{tabular} 
\\ {\bf Services contd\ldots  (Web Application Interface)} & \begin{tabular}{@{}p{.6\textwidth}}

{\bf Operation:} render admin tasks portal \\
{\bf Description:} Render the user interface showing the pending tasks for an admin as well as all pending tasks grouped by respective course events. \\ \\

{\bf Operation:} render submissions view \\
{\bf Description:} Render the user interface containing student submissions to programming assignments.  \\ \\

{\bf Operation:} render code viewer \\
{\bf Description:} Renders the user interface for viewing code history and grading. \\ \\

{\bf Operation:} grade submission \\
{\bf Description:} Sets the grade for a programming assignment submission for the course. \\ \\

{\bf Operation:} render assignment analytics \\
{\bf Description:} Renders the user interface containing analytics for a particular assignment. \\ \\

{\bf Operation:} render admin editor \\
{\bf Description:} Renders the user interface from which admin users can compile, execute and evaluate arbitrary code. The editor is also used to update solution codes to problems. \\ \\
\end{tabular} 
\\ {\bf Services contd\ldots  (Web Application Interface)} & \begin{tabular}{@{}p{.6\textwidth}}

{\bf Operation:} render control panel \\
{\bf Description:} Renders the user interface which is used to modify settings for the tutoring system such as execution delays and compiler flags. \\ \\

{\bf Operation:} render admin settings \\
{\bf Description:} Renders the user interface from where admin users can modify their name and password. \\ \\

{\bf Operation:} render admin pager \\
{\bf Description:} Renders the user interface using which admin users will be able to view message threads created by students. \\ \\
\end{tabular} \\ \hline 
{\bf Protocol} &
All activities using this interface must be preceded by a login. \\ \hline 
{\bf Remarks} &
Exposed as a web service. \\ \hline 
{\bf Issues / Enhancements} &
Error handling not stable. Cleanup required in the codebase.  \\ \hline 
\end{longtable}
\ \\[8mm]

\begin{tabular}{|p{.3\textwidth}|p{.65\textwidth}|} \hline
{\bf Interface} &
Engine Endpoint \\ \hline
{\bf Description} &
Provides access to the Engine services.  \\ \hline
{\bf Services} &
\begin{tabular}{@{}p{.6\textwidth}}
{\bf Operation:} tool() \\
{\bf Description:} Runs a tool specified in the set of parameters sent to the server. Additional parameters may also be passed to this method, as required by it. \\ \\

{\bf Operation:} compile() \\
{\bf Description:} Compiles a program, which is specified as a parameter and returns the result of compilation. This method may also log results, based on the context of the request. \\ \\

{\bf Operation:} execute() \\
{\bf Description:} Executes a program on a test case as specified in the set of parameters, and returns the result of execution. This method may also log the results based on the context of the request. \\ \\

{\bf Operation:} evaluate() \\
{\bf Description:} Evaluates a programming assignment submission on a set of test cases that were assigned to the problem, and returns the result. This method may log results depending on the context in which it was invoked. \\ \\  
\end{tabular}\\ \hline
{\bf Protocol} &
Compilation followed by execution or evaluation. \\ \hline
{\bf Remarks} &
Exposed as a HTTP web service. \\ \hline
{\bf Issues / Enhancements} &
Sequence of requests is to be enforced.  \\ \hline
\end{tabular}

\begin{tabular}{|p{.3\textwidth}|p{.65\textwidth}|} \hline
{\bf Interface} &
Cache \\ \hline
{\bf Description} &
Provides access to the In-Memory cache.\\ \hline
{\bf Services} &
\begin{tabular}{@{}p{.6\textwidth}}
{\bf Operation:} get key \\
{\bf Description:} Fetches the value corresponding to a key from the cache. \\ \\
{\bf Operation:} set key \\
{\bf Description:} Sets a key along with its value, in the cache. \\ \\
\end{tabular} \\ \hline
{\bf Protocol} &
There are no restrictions on accessing this interface.\\ \hline
{\bf Remarks} &
A single interface is exported by a single cache node.\\ \hline
{\bf Issues / Enhancements} &
No issues currently.\\ \hline
\end{tabular}
\ \\[8mm]

\begin{tabular}{|p{.3\textwidth}|p{.65\textwidth}|} \hline
{\bf Interface} &
Database Node \\ \hline
{\bf Description} &
Provides access the database server. \\ \hline
{\bf Services} &
\begin{tabular}{@{}p{.6\textwidth}}
{\bf Operation:} run sql queries \\
{\bf Description:} Runs SQL queries against the database server hosted on that node. \\ \\
\end{tabular} \\ \hline
{\bf Protocol} &  
There are no restrictions on accessing this interface. \\ \hline
{\bf Remarks} &
A database node exports a single interface for it. \\ \hline
{\bf Issues / Enhancements} &
No issues currently.\\ \hline
\end{tabular}
}

\begin{figure}[t]
\centering
\includegraphics[width=\textwidth]{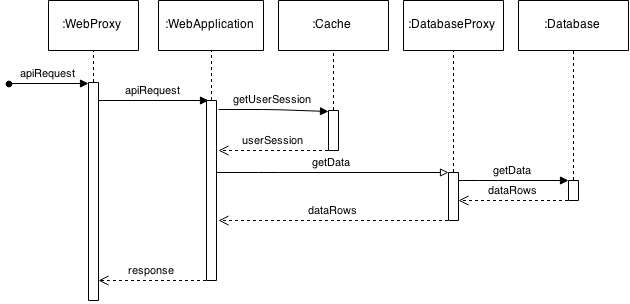}
\caption{Sequence diagram for a Web API Request}
\label{fig:web_api}
\end{figure}

\begin{figure}[ht!]
\centering
\includegraphics[width=\textwidth]{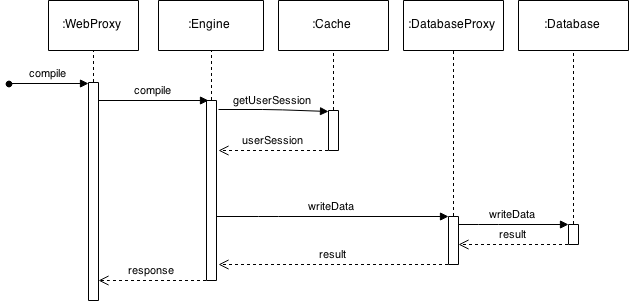}
\caption{Sequence diagram for a compilation request}
\label{fig:compile_request}
\end{figure}

\subsection{Component Interaction Model}

Figure~\ref{fig:web_api} represents the flow of a basic API request from the client. The request first reaches the HTTP proxy and load balancer, which forwards it to one of the web application nodes based on its load balancing algorithm. The web application node, upon receiving it first checks whether the request is authentic or not. It does so by using the cookie information supplied with the request and consulting with the in-memory cache to fetch the corresponding session data. If no such data is present, then the request is not authentic. A request may also not be authentic, if the roles as specified in the session data, do not match with the role required the access the service. Once the request is authenticated, the appropriate modules are invoked and database accesses are made via the database proxy. The database proxy forwards these requests to one of the database nodes based on the load balancing algorithm used. Once the required result is available from the modules, it is sent back to the client. 

Figure~\ref{fig:compile_request} represents the flow of a basic engine request. All engine service requests arrive at the web proxy which forwards them to an engine node. The engine then uses its session manager to look up the cache and verify the authenticity of the request. After verifying the authenticity, the corresponding tools for the request are invoked in the engine and database accesses are made via the database proxy. The final results of execution of required tools are sent as a marshalled JSON string, to the client.

\begin{figure}[t]
\centering
\includegraphics[width=\textwidth]{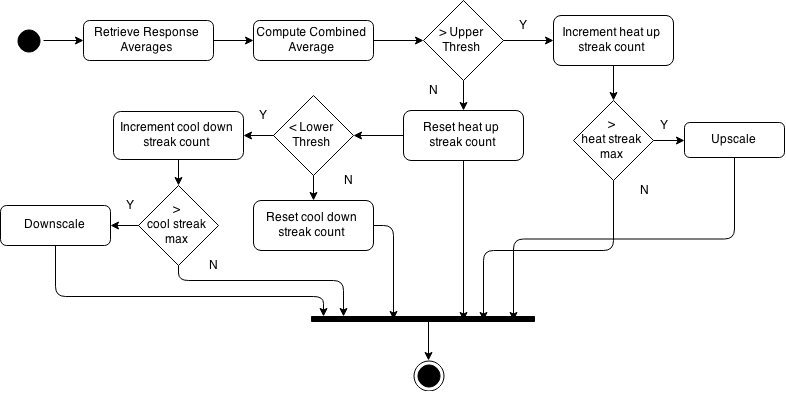}
\caption{Activity diagram of scaling monitor for auto scaling web services}
\label{fig:auto_scaling}
\end{figure}

\subsection{Auto Scaling}
For web application nodes, the response time is logged by the application server. This log is used to compute the average response time over last n requests. This average response time is published to a distributed key-value store at regular intervals. A monitor client for the host machine, which resides on the host machine itself, keeps track of these published response time entries. It first gathers information regarding the running containers on the host machine, and then computes the average over the response times published by each of the individual nodes on the key-value store. This average is used as a metric for the load on the web application servers. Once this value goes high, and stays high for a specific number of intervals, an upscale is triggered. This situation is called heat-up. The number of intervals for which the cumulative average stays high is referred to as the streak length. Once upscale is triggered, a new instance of the web application is spawned and connected to the cluster and load balancer. Similarly, if the average across all the web application nodes falls below a certain threshold and stays as such for a specific streak length, cooldown occurs, and a  downscale is triggered. Downscaling simply removes a running instance from the cluster of web applications. The streak length for heat-up is much smaller than the streak length for cooldown (typically 1/10th). The time taken to provision a new instance is quite small (order of a few seconds). Figure~\ref{fig:auto_scaling} shows the activity diagram for auto scaling web services.

The auto scaling algorithm had been tested by subjecting the system to stress tests. The system was flooded with requests gradually, and the number of nodes increased to a specific value. Similarly, when the flooding was over, the newly spawned nodes were destroyed after some time. Scalability tests were also carried out manually. The response times (a measure of the performance of the system), decreased for the web application and engine subsystems, as more nodes were added to the system. According to the benchmarks, this was somewhat linear. The database also scaled slowly up to a certain number of nodes, till it reached the saturation point. This was understandable, as the clustering scheme used for the database was synchronous replication. In order to scale up the database further, the data model of the system needs to be changed.

\begin{figure}[t]
\centering
\includegraphics[width=\textwidth]{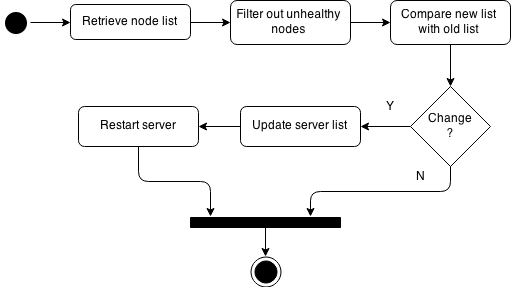}
\caption{Activity diagram of load balancing monitor for dynamic load balancing}
\label{fig:load_balancing}
\end{figure}

\subsection{Dynamic Load Balancing}
Due to the scalable nature of the system, nodes are added and removed from it dynamically. For the load balancer, this requires keeping track of what web service nodes are added to the system and what are removed. The load balancer forwards all HTTP requests to one of the web application or engine servers, based on the load balancing algorithm that it uses. However, in order to forward these requests, it must know the set of servers that are available. One cannot statically assign such servers to its server list as these can change arbitrarily. Thus the load balancer uses the services of the service discovery agent, to acquire information about what nodes are available and healthy. It then uses this information to dynamically update its server list, whenever there is a change. Figure~\ref{fig:load_balancing} shows the activity diagram of load balancing monitor for dynamic load balancing.

    \section{Experience with Prutor}
Prutor has been used to teach the introductory programming course (ESC101) at IIT Kanpur\footnote{\url{http://www.iitk.ac.in/esc101}} since 2014-15 autumn semester. Recently it has been deployed at IIT Bombay\footnote{\url{http://www.cse.iitb.ac.in/~cs101}} (since 2015-16 spring semester), IIT Goa (since 2016-17 autumn semester), and IISER Bhopal (Since 2016-17 autumn semester). The maximum students enrolled for the course at IIT Kanpur has been around 430 and at IIT Bombay around 560. This section summarizes our experiments and experiences.
\begin{itemize}
\item \textbf{Handling Language Diversity}

Prutor has been experimented with several different programming languages including C, C++ and Python. The framework is expected to adapt to any imperative programming language with minor modifications in configuration files. This ability to handle language diversity is a robust feature of Prutor.

\item \textbf{Syntactic Feedback}

The framework was experimented with a couple of syntactic feedback tools which were used to provide real-time feedback to students on syntactical errors performed while writing programs. These two tools differed in the mechanism in which feedback was generated and also the sort of errors that could be captured using them. One of the tools has the capability to generate repairs for the programs.
\begin{itemize}
\item Compiler message rephrasing: This feedback tool reads the messages generated by the underlying compiler and rephrases them into more meaningful an accurate messages, according to some rule defined for such errors messages. These rules are manually entered by the instructor on the basis of the common errors committed by the students. The tool gives a ranking of the types of errors based on the frequency in which they are committed.
\item Clustering algorithms: This feedback tool depends on rules written by the instructor to capture and repair errors. The rules are applied on the source code of the program written by the student. 
\end{itemize}

\item \textbf{Syntactic Repair}

In addition to error detection, error repairing tools have also been integrated into Prutor. A syntactic error repair tool based on rules defined by the instructor, have been integrated into the framework. The tool can run on uncompiled code and detect if there are any errors (according to the defined rules) in the code. It then generates the candidate repairs for it.

\cmt{ 
\item \textbf{Semantic Feedback (Ivan's tool)}

\item Grading (Manual vs Automated)
\item Dashboard
\item Extensions to handle other Programming Languages
\item Graphs and charts comparing 
\begin{itemize}
\item student learning curve across labs
\item rates of various things 
\item TA/feedback rate
\item grade distribution across offerings
\end{itemize}}

\item \textbf{Failures and Recovery}
There has been no data loss in the system till now. Failure of nodes did not affect the integrity or consistency of data in the system. Removal of database nodes from the system did not result in loss of data. This is consistent with the fact that synchronous replication is being used. Thus the system is indeed durable.

There had been several occasions when a number of nodes of the system had crashed, due to bugs in software or other reasons. However, service was not disrupted and the failures were invisible to most of the end users. Only the few who were using the respective node a the time when it crashed, were affected for a brief period of time. This is due to the fault-tolerant design of the system, where multiple redundant nodes are kept to mitigate failures.  

\end{itemize}

    
    \section{Conclusions}
The following features of Prutor make it an attractive choice for any programming language course:
\begin{itemize}
\item Prutor is {\em portable}. It is designed to run on almost any
  flavour of Linux. The nodes can be packed into archives and
  relocated to different physical machines. Only the environment which
  is required to run the nodes must be set up on the physical
  machines. Docker~\cite{docker}, the environment used for deploying
  nodes, can easily be set up on any standard Linux distribution.

\item Prutor is {\em extensible}.  Prutor is a plug-and-play
  system. Any plugin that follows a certain convention can be added to
  Prutor to extend Prutor's functionality. Adding plugins for specific
  purposes such as feedback generation, requires only a few lines of
  configuration and code. Adding a plugin does not require any
  downtime.
  
\item Prutor is easily {\em modifiable}: The existing tools and modules in the system can be modified quite easily by changing a few lines of code. Few tools such as compilers can be changed by simply changing the compilation configurations. Features can also be enabled and disabled via the admin interface using simple clicks.
  
\item Prutor is {\em scalable}: New nodes can be added and removed to and from the system without any downtime. New physical machines can also be added to the system, without affecting the uptime. The system has the capability to auto scale nodes within a physical machine, in the case when the machine is used for multiple purposes.
  
\item {\em Usability}: The system is meant to be used by novice computer users to expert programmers. The user interfaces are modular and intuitive. 
\item {\em Durability}: The data nodes on the system are replicated, thus making it immune to data loss. 

\item {\em Fault-Tolerance}: All nodes on the system are replicated, thus making the system highly available. Failure of nodes does not result in downtime due to the presence of redundant replicas. The proxies also operate in high availability master-slave mode, thus adding a new layer of tolerance at the physical level.
The physical machines have proxies that listen on a single floating IP with a master-slave configuration. Thus, in the event that one of the physical machines go down, the other machine’s proxy takes over. This principle supports fault tolerance. 
\end{itemize}

The only major constraint of the Prutor system is that it has to run
on a Linux environment. However any bare minimal Linux distribution
which supports Docker will suffice.

Docker is used as the virtualization environment for the system. This
results in the system being easily portable. However, Docker’s
networking does not currently support multiple hosts. To address this
issue, Weave has been used to create a common Docker network for
containers residing on multiple hosts. Currently there are no issues
to this approach, except for the fact that Docker itself does not
recognize Weave. There is no way to obtain the information about what
IP address is assigned to a container using the Weave network, via
Docker. This has to be done manually by entering the container.

This system is architected in such a way that it can handle a load of
about 500 students accessing the system simultaneously. It is easily
deployable, owing to the use of Docker containers. This enables it to
be deployed across multiple machines with ease and not requiring much
expertise. Thus instructors of various programming courses may install
this system on the servers available at their institutions. This
system is also designed to run off any standard machine. This implies
that it can even be run off a laptop machine. Most of the features
provided by the architecture of this system reduces the amount of work
required by system administrators in maintaining it. The highly
modular nature of the codebase provides a great framework for creating
variants of this tutoring system.

Multiple analytics tools have been integrated into the system, for
research purposes. This has proved the system to be extensible. These
tools were integrated by simply adding a new module into the Engine
subsystem. The web services provided by this module were then
registered in the routes of the Engine.

The system had been modified to work with the Python interpreter as well as some basic Prolog. The required changes to the system took around 2 days with a single person’s effort. In general, in order to allow the system to work for a new imperative programming language, the only modification required is to change the configuration files. The system can also be moulded to include other tools as well, such as feedback tools. This makes the system highly modifiable.

    \section{Acknowledegements}
IIT Kanpur for the facilities provided.
    
    \bibliographystyle{abbrv}
    \bibliography{references}

\begin{thebibliography}{1}

\bibitem{consul}
Consul, {https://www.consul.io/}.

\bibitem{coursera}
Coursera: Free online classes for everyone.
\newblock http://www.coursera.org.

\bibitem{edx}
Edx: New online learning initiative of {H}arvard university and {MIT}.
\newblock https://www.edx.org/.

\bibitem{khan-academy}
Khan academy.
\newblock http://www.khanacademy.org/.

\bibitem{docker}
D.~Merkel.
\newblock Docker: Lightweight linux containers for consistent development and
  deployment.
\newblock {\em Linux J.}, 2014(239), Mar. 2014.

\bibitem{nptel}
National programme on technology enhanced learning.
\newblock http://nptel.iitm.ac.in.

\bibitem{udacity}
Udacity: Free online university classes for everyone.
\newblock http://www.udacity.com/.

\end{thebibliography}
    \appendix
    \section{Domain Lexicon}
\begin{figure}[t]
  \centering
  \includegraphics[width=\textwidth]{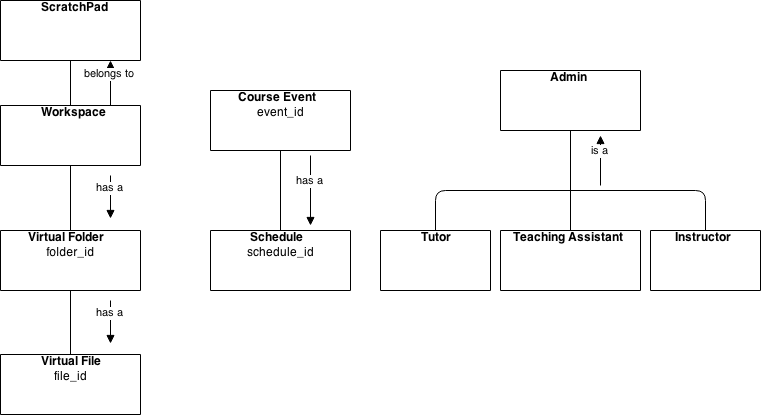}
  \caption{Lexicon diagram}
  \label{fig:lexicon}
  
\end{figure}
\begin{itemize}
\item{\bf Engine:} The component of the system which is responsible for tasks such as compilation, execution, evaluation and feedback generation.

\item{\bf Virtual File:} An identifier for a student program, created on the system, which appears as a file to the student. Actual files are not created on the disk.

\item{\bf Virtual Folder:} An identifier for a collection of programs, created on the system, which appears as a folder to the student. Actual folders are not created on the disk.

\item{\bf Workspace:} A collection of virtual files and folders, used by a student for programming purposes.

\item{\bf ScratchPad:} A web based IDE interface, wherein a student can create virtual files and folders as a part of her workspace. Students can arbitrary write programs in such files using a provided editor window, in the programming language assigned for the course. These programs can be compiled and their results can be viewed in the form of annotations in the gutter area of the editor window. Compiler messages can be viewed in a virtual console located within the same user interface. Further, the student may execute her compiled program on arbitrary test cases and view the results in an output window, within the same user interface. Execution results such as runtime errors or time limit exceeded errors can also be viewed. 

\item{\bf CodeBook:} A portal where students can view submitted solutions to programming assignments along with the corresponding problem statements for course events that were held previously. Grading information for the submitted assignments can also be viewed in the same interface. The portal also contains solutions to practice problems attempted by the student, along with the corresponding problem statements.

\item{\bf Course Event:} Any event which is conducted during the course offering, and includes programming assignments to be solved by students. Such events are expected to have a set of programming problems assigned to each student who is enrolled for the course. The events may span multiple days or have multiple schedules. Lab assignments, examinations and quizzes are examples.

\item{\bf Schedule:} A time period which corresponds to a course event, during which, members of that schedule may solve the programming problems assigned to them. All programming problems assigned to a particular student for a course event are only accessible to her during the time span of the schedule. 

\item{\bf Practice Problems:} A set of problems, which are not associated with any course event. They are accessible at all times, and can be solved by the student at all times. They are intended to provide practice to the students in solving programming problems.

\item{\bf PracticeArena:} A portal where a student can access a collection of programming problems for practice. Students may view these problems and navigate to the editor interface, using which they can solve the problem.

\item{\bf Admin:} A teaching assistant, tutor or an instructor of the course.

\item{\bf Teaching Assistant:} A student who helps the instructor of the course in carrying out activities such as grading, invigilation and helping students solve programming problems.

\item{\bf Tutor:} A student or a professor who has the responsibilities of setting questions for various course events and deciding the policies of the course events.

\item{\bf Instructor:} A professor who conducts the course. He is responsible for deciding the course structure and grading policies for the course.

\item{\bf GradeCard:} A table containing scores awarded by teaching assistants, tutors and instructors for the programming problems that were solved by the student for various course events. These scores are grouped by the course events attended by the student.

\item{\bf Pager:} A publish-subscribe based messaging system wherein a student can create a message thread when she requires help in solving programming problems or in addressing technical difficulties, while solving the problems. Instructors, tutors and teaching assistants are able to view these messages and respond to them in real time. Only students can start a message thread, while others can only respond to them.

\item{\bf DataViz:} A portal where data collected using the system can be tabulated in arbitrary formats or visualizations may be created from the data.

\item{\bf ControlPanel:} A portal from where administrators can tune the settings of the Engine. Settings include time delays in compilation and execution along with execution quotas, compiler flags and tool active state.

\item{\bf CodeViewer:} An user interface where the code history corresponding to a student submission for a programming assignment can be viewed. Information regarding the category of a code save along with any results of compilation or execution can be viewed for the entire code history. Code saves may also be deleted from the system using this interface. Further, admins can grade the programming assignments using this interface. 
\end{itemize}

    \section{Process Views}

\begin{figure}[t]
  \centering
  \includegraphics[width=\textwidth]{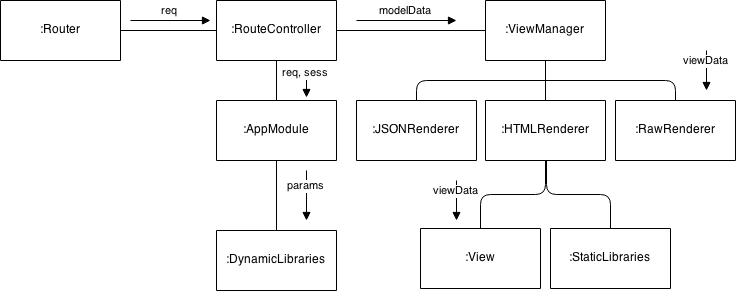}
  \caption{Collaboration diagram for web application subsystem}
  \label{fig:web_app_collab}
\end{figure}

\begin{figure}[t]
  \centering
  \includegraphics[width=\textwidth]{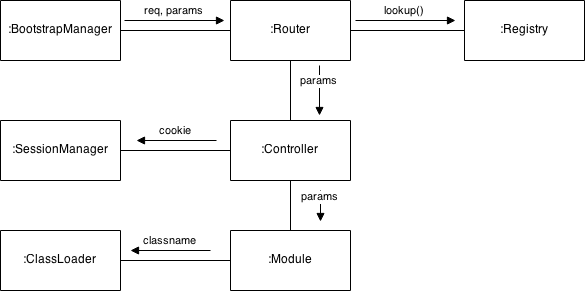}
  \caption{Collaboration diagram for Engine subsystem}
  \label{fig:engine_collab}
\end{figure}

\begin{figure}[t]
  \centering
  \includegraphics[width=\textwidth]{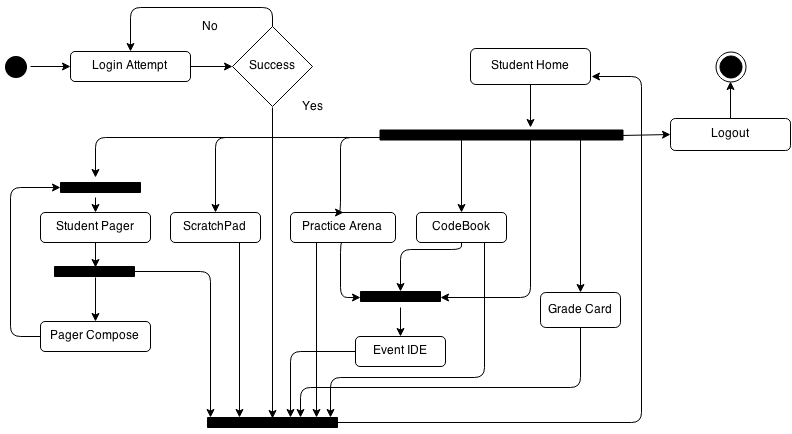}
  \caption{Normal student activity}
  \label{fig:student_act}
\end{figure}

\begin{figure}[t]
  \centering
  \includegraphics[width=\textwidth]{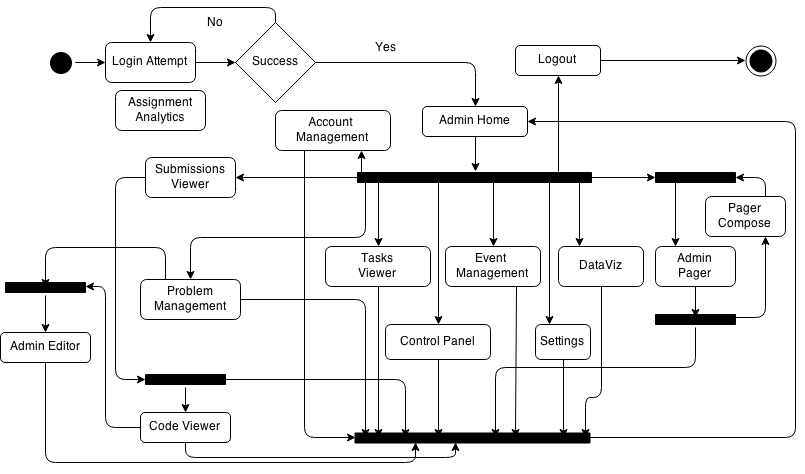}
  \caption{Normal admin activity}
  \label{fig:admin_act}
\end{figure}

\begin{figure}[t]
  \centering
  \includegraphics[width=\textwidth]{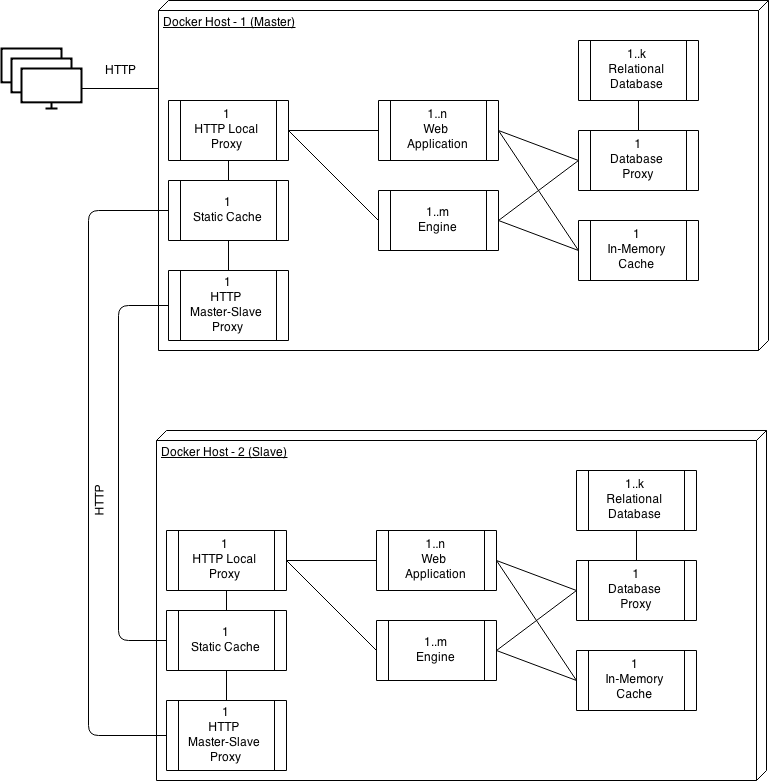}
  \caption{Deployment view of the tutoring system}
  \label{fig:deploy}
\end{figure}

\end{document}